\documentstyle[epsfig]{mn2e}

\def\be{\begin{equation}}
\def\ee{\end{equation}}
\def\bea{\begin{eqnarray}}
\def\eea{\end{eqnarray}}

\title[]{Late internal shock model for bright X-ray flares in
Gamma-ray Burst afterglows and GRB 011121}

\author[]{Y. Z. Fan$^{1,2,3 \star}$ and
 D. M. Wei$^{1,2}$
\thanks{E-mail: yzfan@pmo.ac.cn(YZF);
dmwei@pmo.ac.cn(DMW)} \\
$^1${\sl Purple Mountain Observatory, Chinese Academy of
Science, Nanjing 210008, China}\\
$^2${\sl National Astronomical
Observatories, Chinese Academy of Sciences, Beijing 100012,
China}\\
$^3${\sl Dept. of Physics, University of Nevada, Las Vegas, NV
89154, USA.}}

\date{Accepted ......  Received ......; in original form ......}

\begin{document}

\maketitle
\begin{abstract}
We explore two possible models which might give rise to bright
X-ray flares in GRBs afterglows. One is an external
forward-reverse shock model, in which the shock parameters of
forward/reverse shocks are taken to be quite different. The other
is a so called ``late internal shock model", which requires a
refreshed unsteady relativistic outflow generated after the prompt
$\gamma-$ray emission. In the forward-reverse shock model, after
the time $t_\times$ at which the RS crosses the ejecta, the flux
declines more slowly than $(t_\oplus/t_\times)^{-(2+\beta)}$,
where $t_\oplus$ denotes the observer's time and $\beta$ is the
spectral index of the X-ray emission. In the ``late internal shock
model", decaying slopes much steeper than $(t_\oplus/t_{\rm e,
\oplus})^{-(2+\beta)}$ are possible if the central engine shuts
down after $t_{\rm e, \oplus}$ and the observed variability
timescale of the X-ray flare is much shorter than $t_{\rm e,
\oplus}$.

The sharp decline of the X-ray flares detected in GRB 011121, XRF 050406, GRB 050502b, and GRB 050730  rules  out
the external forward-reverse shock model directly and favors  the ``late internal shock
model". These X-ray flares could thus hint that the central engine
operates again and a new unsteady relativistic outflow is generated just a few minutes after the intrinsic hard burst.
\end{abstract}


\section{Introduction}
\label{sec:XRB1}
 GRB 011121 was simultaneously detected by {\it BeppoSAX} GRBM and
 WFC (Piro 2001), and the fluence in the
2-700 keV range corresponds to an isotropic energy of $2.8\times 10^{52}$ergs at the redshift of $z=0.36$ (Infante et al.
2001). This burst was born in a stellar wind (Price et al. 2002; Greiner et al. 2003) and a supernova bump was
detected in the late optical afterglow (Bloom et al. 2002; Garnavich et al. 2003). Its very early X-ray
light curve, which has not been published until quite recently,  is characterized by the presence of two flares
(Prio 2005, hereafter P05). In the first one, which is also the strongest of the two, the observed flux $F$ rises and decays very steeply: $F\propto t_\oplus^{10}$ for $239~{\rm s}<t_\oplus<270~{\rm s}$ and $F\propto t_\oplus^{-7}$
for $270~{\rm s}<t_\oplus<400~{\rm s}$, where $t_\oplus$ is the observer's time\footnote{During our revision,
 bright X-ray flares peaking a few
minutes after XRF 050406, GRB 050502b and GRB 050730 have been
reported (Burrows et al. 2005; Starling et al. 2005). Sharp rise
and fall are also evident in these events.}. Such a peculiar flare
in the early X-ray lightcurve of GRBs has not been predicted
before. P05 suggested that the X-ray flare represents the
beginning of the afterglow.

In this Letter we explore two alternative models which might give
rise to very early X-ray flare in GRB afterglows
(\S\ref{sec:XRB2}): a forward-reverse shock model
(\S{\ref{sec:XRB21}}) and  a ``late internal shock model"
(\S{\ref{sec:XRB22}}). We compare the available data with the
predictions of those models in \S{\ref{sec:XRB3}} and summarize
our results in \S{\ref{sec:XRB4}},  with some discussion.

\section{Possible models}
\label{sec:XRB2}

\subsection{The external forward-reverse shock model}
\label{sec:XRB21}

The external forward-reverse shock model has been widely accepted on interpreting the early IR/optical flashes of GRB 990123,
GRB 021211 and GRB 041219a (For observations, see: Akerlof et al. 1999; Fox et al. 2003; Li et al. 2003; Blake et al.
2005. For theoretical modeling, see: Sari \& Piran 1999; M\'esz\'aros \& Rees 1999; Wei 2003; Kumar \& Panaitescu 2003;
Fan, Zhang \& Wei
 2005). Synchrotron radiation from the reverse shock (RS) and the forward shock (FS) usually peaks in the infrared-to-optical and ultraviolet-to-soft X-ray bands, respectively. Thus, the RS emission component is not dominant in the X-ray band. The
synchrotron self-Compoton (SSC) scattering effect of the RS radiation has also been considered by different authors, but no strong
X-ray emission is found to be expected (Wang, Dai \& Lu 2001) except in some carefully balanced conditions (Kobayashi et al.
2005).

In most of previous works, the fractions of FS energy
given to electrons, $\epsilon_{\rm e}$, and to magnetic field, $\epsilon_{\rm B}$, were assumed to be the same as the
corresponding fractions in the RS. However, this may not necessary be the case. Fan et al. (2002) performed a detailed fit to the optical flash of GRB 990123 data and obtained
 $\epsilon_{\rm e} ^{\rm r}=4.7\epsilon_{\rm
e}^{\rm f}$ and $\epsilon_{\rm B}^{\rm r}=400\epsilon_{\rm B} ^{\rm f}$, where the superscripts
``r'' and ``f'' represent RS and FS, respectively. Similar results were obtained by Zhang, Kobayashi \&
M\'esz\'aros (2003), Kumar \& Panaitescu (2003), Panaitescu \& Kumar (2004), McMahon, Kumar
 \& Panaitescu (2004), and Fan et al. (2005). In this section, we study the RS/FS emission in X-ray band by adopting
different shock parameters. We focus on the thin shell case (i.e., the RS is
sub-relativistic, see Kobayashi [2000]), in which the RS emission is well separated from the prompt $\gamma-$ray
emission.

{\bf ISM model.} In the thin shell case, the observer's time at which RS crosses the ejecta can be estimated by (e.g., Fan et al. 2005)
\begin{equation}
t_{\times}\approx 128{\rm s}~({1+z\over 2})E_{\rm iso,53}^{1/3}n_0^{-1/3}\eta_{2.3}^{-8/3},
\label{Eq:t_times}
\end{equation}
where $E_{\rm iso}$ is the isotropic energy of the outflow, $n$ is
 the typical number density of ISM, and $\eta$ is the initial Lorentz factor of the outflow.
 The convention $Q_{\rm y}=Q/10^{\rm y}$ has been adopted in cgs units throughout the text except for some special notations.

In the standard afterglow model of a fireball interacting with a
constant density medium (e.g., Sari, Piran \& Narayan 1998), the
cooling frequency $\nu_{\rm c,\oplus}^{\rm f}$, the typical
synchrotron frequency $\nu_{\rm m,\oplus}^{\rm f}$, and the
maximum spectral flux $F_{\rm \nu,max}^{\rm f}$ read: $\nu_{\rm
c,\oplus}^{\rm f}=4.4 \times 10^{17}~{\rm Hz}~E_{\rm
iso,53}^{-1/2} \epsilon_{\rm B,-3}^{-3/2} n_0^{-1} {t}_{\rm
d,-3}^{-1/2} ({2\over 1+z})(1+Y^{\rm f})^{-2}$, $\nu_{\rm
m,\oplus}^{\rm f} = 4.4 \times 10^{15}~{\rm Hz}~E_{\rm iso,
53}^{1\over 2} \epsilon_{\rm B,-3}^{1\over 2}$ $\epsilon_{\rm
e,-1}^2 {t}_{\rm d,-3}^{-{3\over 2}} C_{\rm p}^2 ({2\over 1+z})$,
and $F_{\rm \nu,max}^{\rm f}=2.6 ~{\rm mJy}~E_{\rm iso,53}
\epsilon_{\rm B,-3}^{1/2} n_0^{1/2} D_{\rm L,28.34}^{-2}
({1+z\over 2})$,  where $C_{\rm p}=13(p-2)/[3(p-1)]$, $p\sim2.3$
is the typical power-law distribution index of the electrons
accelerated by FS, $Y^{\rm f}\simeq [-1+\sqrt{1+4x^{\rm f}
\epsilon_{\rm e}^{\rm f}/\epsilon_{\rm B}^{\rm f}}]/2$ is the
Compton parameter, $x^{\rm f}\approx {\rm min}\{1,~(\nu_{\rm m}^
{\rm f}/\nu_{\rm c}^{\rm f})^{\rm (p-2)/2}\}$  (Sari \& Esin
2001), and $D_{\rm L}$ is the luminosity distance for
$(\Omega_{\rm M}, ~ \Omega_{\Lambda}, ~h)=(0.3,~0.7,~0.71)$.
Hereafter $t=t_\oplus/(1+z)$, and $t_{\rm d}$ is in unit of days.

Following Zhang et al. (2003), we take $\epsilon_{\rm e}^{\rm r}=
{\cal R}_{\rm e}\epsilon_{\rm e}^{\rm f}$ and $\epsilon_{\rm
B}^{\rm r}={\cal R} _{\rm B}^2 \epsilon_{\rm B}^{\rm f}$. At
$t_\times$, the RS emission satisfies [See also Fan et al. (2005),
note that a novel effect taken into account here is the inverse
Compton cooling of the electrons] \bea \nu_{\rm m,\oplus}^{\rm
r}(t_{\rm \times})&=&{\cal R}_{\rm B}[{\cal R}_{\rm
e}(\gamma_{34,\times}-1)]^2
\nu_{\rm m,\oplus}^{\rm f}(t_{\rm \times})/(\Gamma_{\times}-1)^2,\\
\nu_{\rm c,\oplus}^{\rm r}(t_\times)&\approx & {\cal R}_{\rm B}^{-3}[(1+Y^{\rm f})
/(1+Y^{\rm r})]^2\nu_{\rm c,\oplus}^{\rm
f}(t_\times), \\
F_{\rm \nu, max}^{\rm r}(t_{\rm \times})&\approx & \eta {\cal
R}_{\rm B} F_{\rm \nu, max}^{\rm f}(t_{\rm \times}).
\eea
where $\gamma_{34,\times}\approx (\eta/\Gamma_\times+\Gamma_\times/\eta)/2$ is the
Lorentz factor of the shocked ejecta relative to the initial one, $\Gamma_\times$
is the bulk Lorentz factor of the shocked ejecta at $t_\times$,
 $Y^{\rm r}\simeq [-1+\sqrt{1+4x^{\rm r}{\cal R}_{\rm e}\epsilon_{\rm e}^{\rm f}/({\cal R}
_{\rm B}^2\epsilon_{\rm B}^{\rm f})}]/2$ is the Compton parameter and $x^{\rm r}\approx
 {\rm min}\{1,~(\nu_{\rm m}^{\rm r}/\nu_{\rm c}^{\rm r})^{\rm (p-2)/2}\}$.

If both $\nu_{\rm c,\oplus}^{\rm r}$ and $\nu_{\rm c,\oplus}^{\rm
f}$ are below the observed frequency $\nu_\oplus$, the detected
flux of RS and FS emission are
\be
F_{\rm \nu_\oplus}^{\rm
r}(t_\times)\simeq F_{\rm \nu,max}^{\rm r}(t_\times) [\nu_{\rm
m,\oplus }^{\rm r}(t_\times)]^{\rm (p-1)/2}[\nu_{\rm c,\oplus}^{\rm r}(t_\times)]^{\rm 1/2}
\nu_\oplus^{\rm -p/2},
\ee
\be
F_{\rm \nu_\oplus}^{\rm f}(t_\times)\simeq F_{\rm \nu,max}^{\rm f}(t_\times)
[\nu_{\rm m,\oplus }^{\rm f}(t_\times)]^{\rm (p-1)/2}[\nu_{\rm c,\oplus }^{\rm f}(t_\times)]^{\rm 1/2}
\nu_\oplus^{\rm -p/2}.
\ee
\be
{F_{\rm \nu_\oplus}^{\rm r}(t_\times) \over F_{\rm \nu_\oplus}^{\rm f}(t_\times)}\approx
\eta {\cal R}_{\rm B}^{\rm p-2\over 2}{\cal R}_{\rm e}^{\rm p-1}
({\gamma_{34,\times}-1 \over \Gamma_\times-1})^{\rm p-1}({1+Y^{\rm f}\over 1+Y^{\rm r}}).
\label{Eq:ratio1}
\ee

\begin{figure}
\begin{picture}(0,300)
\put(0,0){\includegraphics{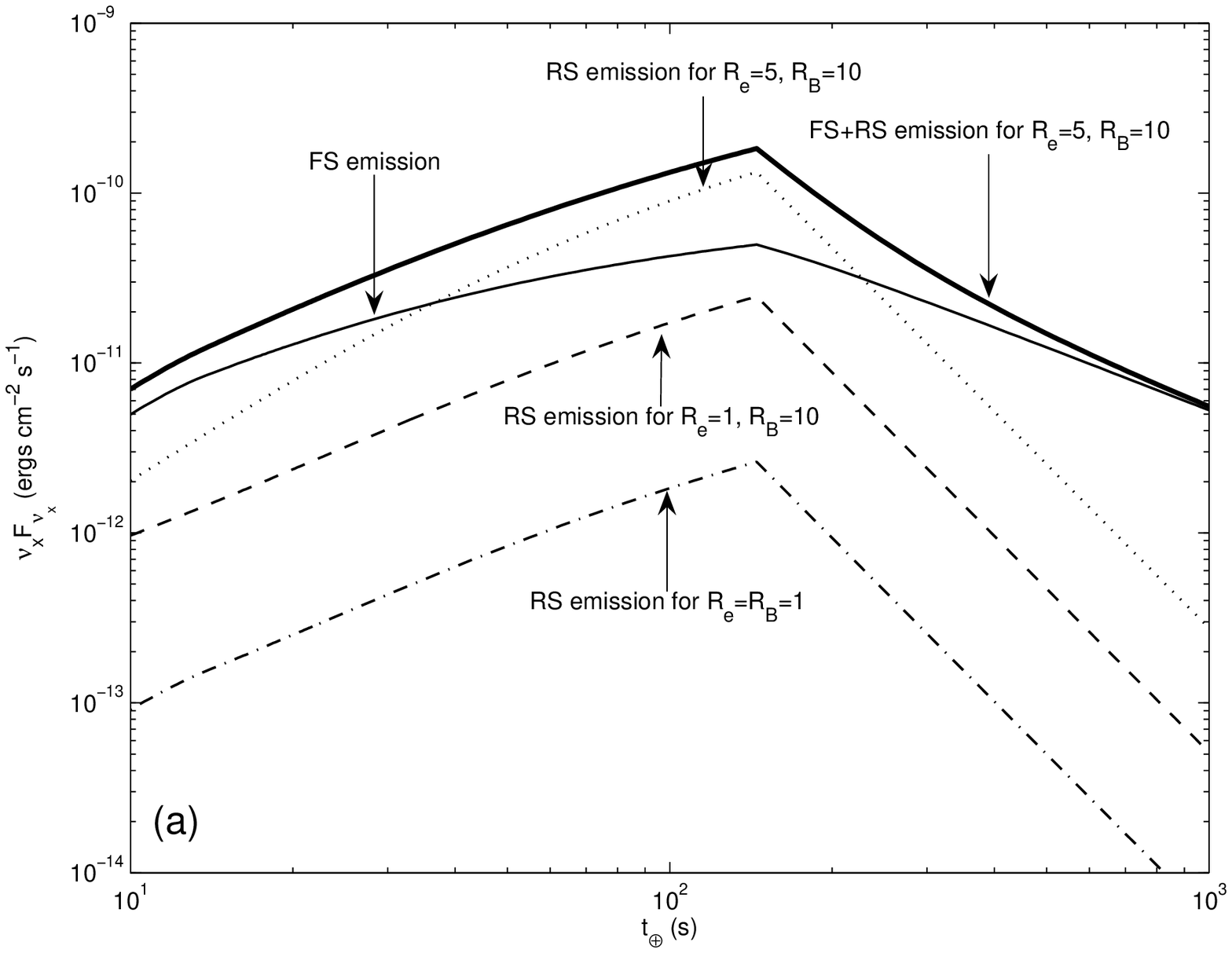}}
\put(0,0){\includegraphics{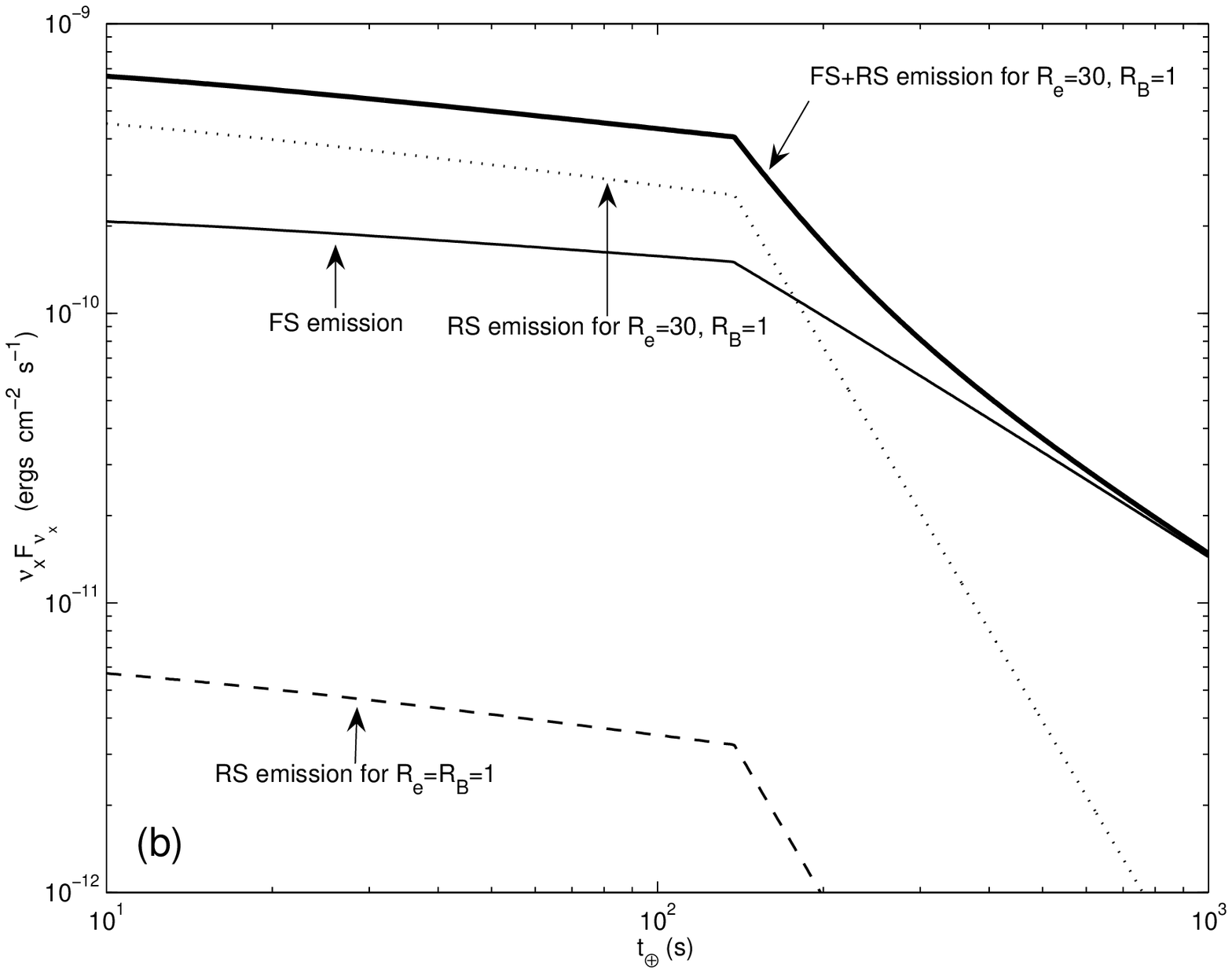}}
\end{picture}
\caption{The very early X-ray ($\nu_{\rm x}=2.42\times 10^{17}{\rm Hz}$)
sample lightcurves.  For the reverse shock emission component, ${\cal R}_{\rm e}$ and ${\cal R}_{\rm B}$ have been marked in the figure.
(a) The ISM case, the parameters are taken as
 $n=1~{\rm cm^{-3}}$, $E_{\rm iso,53}=1$, $z=1$, $\Delta=6\times 10^{11}$ cm,
$\epsilon_{\rm e}^{\rm f}=0.1$, $\epsilon_{\rm B}^{\rm f}=0.001$, $\eta=200$, and $p=2.4$.
 (b) The wind case, following P05, we take $A_*=0.003$, $\epsilon_{\rm e}^{\rm f}=0.01$,
 $\epsilon_{\rm B}^{\rm f}=0.5$, $E_{\rm iso,53}=0.28$, $p=2.5$, and $z=0.36$; In addition, we assume $\eta=200$ and
$\Delta=3.0\times 10^{12}$cm.}
\label{fig:FRX}
\end{figure}

Taking $p=2.3$ (such a choice is to match the observed slope
of the X-ray flare $\beta=p/2=1.15$, see P05), $\Gamma_\times \approx \eta/2\sim 100$, $\epsilon^{\rm f}_{\rm B,-3}=1$, $\epsilon^{\rm f}_{\rm e,-1}=1$,
 ${\cal R}_{\rm B}=10$
and ${\cal R}_{\rm e}=5$, we have $x^{\rm f}\approx 1$, $Y^{\rm f}\approx (\epsilon_{\rm e}/\epsilon_{\rm B})^{1/2}=10$,
 $x^{\rm r}\approx 0.6$, $Y^{\rm r}\approx 1.2$, and $(1+Y^{\rm f})/(1+Y^{\rm r})\gg 1$, i.e., we have a larger  contrast
 $F_{\rm \nu_\oplus}^{\rm r}(t_\times)/F_{\rm \nu_\oplus}^{\rm f}(t_\times)$ when the inverse Compton effect has been
taken into account.
With equation (\ref{Eq:ratio1}),
we have $F_{\rm \nu_\oplus}^{\rm r}(t_\times)/F_{\rm \nu_\oplus}
^{\rm f}(t_\times)\approx 5$, i.e.,  {\it in the X-ray band, the RS emission component is dominant}.
  For $t_\oplus>t_{\times}$, the RS emission declines as $(t_\oplus/t_\times)^{-(2+p/2)}$ because of the curvature effect
 (e.g., Kumar \& Panaitescu 2000, hereafter KP00)\footnote{To derive the curvature effect,  two assumptions
are made (KP00). One is that the Lorentz factor of the outflow is nearly a constant. The other is that
the observer frequency should be above the cooling frequency of the emission. As far as the reverse shock emission mentioned here,
these assumptions are satisfied.  So we take $(t_\oplus/t_\times)^{\rm -(2+p/2)}$ to describe the decline, which has been verified by
 the detailed numerical calculation (Fan, Wei \& Wang 2004; see also Fig. \ref{fig:FRX} of this work).}. The FS emission declines as $t_\oplus^{\rm (2-3p)/4}$ (e.g., Sari et al. 1998), so
 the X-ray flare lasts $\sim [F_{\rm
\nu_\oplus}^{\rm r}(t_\times)/F_{\rm \nu_\oplus}^{\rm
f}(t_\times)]^{\rm 4/(10-p)}t_{\times} \sim 300~{\rm s}$. Taking $z=0.36$,  $Q_{\rm
y}=1$, and $\nu_{\rm x}=2.4\times 10^{17}$ Hz, we have $\nu_{\rm x}
F_{\nu_{\rm x}}^{\rm f}\sim 10^{-9} ~ {\rm
ergs~cm^{-2}~s^{-1}}~[(1+z)/1.36]D_{\rm L,27.7}^{-2}$. The peak flux of the X-ray flare is $\simeq
\nu_{\rm x}[F_{\nu_{\rm x}}^{\rm f} + F_{\nu_{\rm x}}^{\rm r}]\sim
5\times 10^{-9}~{\rm ergs~cm^{-2}~s^{-1}}~[(1+z)/1.36]D_{\rm L,27.7}^{-2}$, which is consistent with the observation of GRB 011121
($\sim 2.4\times 10^{-9}~{\rm ergs~cm^{-2}~s^{-1}}$, see P05). However, the accompanying optical flash is very bright. With the typical parameters taken here, the V band flux is $\sim 5$ Jy.

{\bf Wind model.} GRB 011121 was born in a stellar wind.   The best fit parameters are
 $p=2.5$, $E_{\rm iso,52}=2.8$, $A_* \sim 0.003$, $\epsilon_{\rm e}^{\rm f} \sim 0.01$, and
$\epsilon_{\rm B}^{\rm f} \sim 0.5$ (P05). It is straightforward to show that with proper choice of ${\cal R}_{\rm e}$
and ${\cal R}_{\rm B}$, at $t_\times$, the RS emission may be dominant in the soft X-ray band.

{\bf Numerical results.} Following Fan et al. (2005), the FS-RS emission (in the X-ray band) has been calculated
numerically.  In the ISM case (see Fig.~\ref{fig:FRX}(a)), the parameters are taken as $E_{\rm iso,53}=1$, $p=2.4$,
 $\epsilon_{\rm e}^{\rm f}=0.1$, $\epsilon_{\rm B}^{\rm f}=0.001$, $n=1~{\rm cm^{-3}}$, and the initial width
of the outflow $\Delta= 6\times 10^{11}$ cm. In
the wind case (see Fig.~\ref{fig:FRX}(b)), we take the best fit parameters presented in P05.

As shown in  Fig.~\ref{fig:FRX}(a), in the ISM case, there comes an X-ray flare dominated by the RS emission only
when both ${\cal R}_{\rm B}$ and ${\cal R}_{\rm e}$ are much
larger than unity. In the wind case,
 with proper ${\cal R}_{\rm B}$ and ${\cal R}_{\rm e}$, the RS emission may be dominant in the soft X-ray band,
too. But there is no flare expected since both the FS and RS emission components decrease
continually even at very early time (see also Zou, Wu \& Dai 2005). So the FS-RS scenario is unable to account for the X-ray
flare detected in GRB 011121. Moreover, in the general framework of accounting for X-ray flares in GRB afterglows,
the FS-RS model is further disfavored for the fact that, even in an ISM scenario with parameters suitable for a RS
flare arising above the FS emission, the predicted temporal decay is too shallow compared to the steep decay of X-ray flares
observed so far (see Fig.~\ref{fig:FRX}).

\subsection{Late internal shock model}
\label{sec:XRB22}
In the standard fireball model of GRBs, the $\gamma-$ray emission is powered by internal shocks,
whose duration depends on the active time of the central engine. However, the variability of some GRB afterglows
implies that the activity of the GRB central engine may last much longer than the duration of the prompt emission recorded
by $\gamma-$ray monitors (e.g.,  Dai \& Lu 1998; Granot, Nakar \& Piran
2003; Ioka, Kobayashi \& Zhang 2005).  In addition, it has been proposed that the Fe line observed in some GRB X-ray afterglows could be attributed to a prolonged activity of the central engine (Rees \&
M\'esz\'aros 2000; Gao \& Wei 2005).

A possible mechanism for the re-activity of the central engine could be as follows. During the accretion phase which powers
the prompt $\gamma-$ray emission, a fraction of the material constituting the massive progenitor could possibly be pulled out;
the central engine could thus be restarted at late times by the fall back of part of this material onto
the central collapsar remnant (King et al. 2005).

Here we assume that the central engine restarts a few minutes
after the prompt $\gamma-$ray emission, powering a  new unsteady
relativistic outflow. We suppose that the Lorentz factor of the
ejected material can be highly variable, setting $\Gamma_{\rm
s}\sim $ 10 and $\Gamma_{\rm f}\sim $ 100, as the typical Lorentz
factors of the slow and fast shells, respectively. The masses of
the slow and fast shells are taken as $m_{\rm f}\simeq m_{\rm s}$.
When an inner fast shell catches up with an outer slow shell at a
radius $\sim 2\Gamma_{\rm s}^2 c \delta t_\oplus/(1+z)$ (where
$\delta t_\oplus$ is the observed typical variability timescale of
the X-ray flares), internal shocks are generated. The Lorentz
factor of the merged shell is $\Gamma \approx \sqrt{\Gamma_{\rm f}
\Gamma_{\rm s}}$ (e.g., Piran 1999), and the Lorentz factor of the
internal shocks can be estimated by $\Gamma_{\rm sh}\approx
(\sqrt{\Gamma_{\rm f}/\Gamma_{\rm s}}+\sqrt{\Gamma_{\rm
s}/\Gamma_{\rm f}})/2$.  We call this re-generated internal shocks
as the ``late internal shocks".

\subsubsection{Physical parameters}
The internal shock model has been discussed by many authors (e.g., Paczy\'nski \& Xu 1994; Rees \& M\'esz\'aros 1994;
Daigne \& Mochkovitch 1998, 2000; Piran 1999; Dai \& Lu 2002). Generally speaking, to calculate the synchrotron emission
of internal shocks, the
following parameters are involved:   the outflow
luminosity $L_{\rm m}$, $\delta t_\oplus$,  $\Gamma$,
$\Gamma_{\rm sh}$, and of course the shock parameters $\epsilon_{\rm e}$ and $\epsilon_{\rm B}$.
For typical GRBs, ($L_{\rm m}, ~\delta t_\oplus,~\Gamma,~\Gamma_{\rm sh},~\epsilon_{\rm e},~\epsilon_{\rm B}$)
are taken to be $\sim (10^{52}~{\rm ergs~s^{-1}}, ~0.001-0.01~{\rm s},~100-1000,~{\rm a~few},~0.5,~0.01-0.1)$,
 respectively (e.g., Dai \& Lu 2002).

The time averaged
isotropic luminosity (2-700 keV) of the X-ray re-bursting of GRB 011121 is
 $ L_{\rm x}\sim 6 \times 10^{48}{\rm ergs~s^{-1}}$ (P05).
So we normalize our expression by taking $L_{\rm m}\sim 5\times 10^{49}~L_{\rm x,49}(\epsilon/0.2)^{-1}~
{\rm ergs ~ s^{-1}}$, where $\epsilon$ is the efficiency factor
of the X-ray flare.  We take $\epsilon \sim 0.2$, as found in typical GRBs/XRFs (Lloyd-Ronning
\& Zhang 2004). Such small $L_{\rm m}$ (comparing with the GRB case) implies that
the fallback accretion rate is just $\sim 0.001-0.01$ times that of the prompt accretion, if the efficiency factor of converting
the accretion energy into the kinetic energy of the outflow is nearly a constant (MacFadyen, Woosley \& Heger 2001).

The $\delta t_\oplus$ measured in X-ray flares is significantly longer than that measured in the intrinsic hard burst
(Burrows et al. 2005). In this Letter, we take $\delta t_\oplus\sim $ 10 s. The spectra
of the X-ray flares detected so far are all nonthermal. The optical thin
condition implies a lower limit on the bulk Lorentz factor of the merged shell (e.g., Rees \& M\'esz\'aros 1994)
\be
\Gamma > 11~L_{\rm m,49.7}^{1/5}[1.36/(1+z)]^{-1/5}\delta t_{\oplus,1}^{-1/5},
\ee
The typical radius of the late internal shock is
$R_{\rm int} \approx 2 \Gamma^2 c \delta t_\oplus/(1+z)>5\times 10^{13}~{\rm cm}~L_{\rm m,49.7}^{2/5}[1.36/(1+z)]^{3/5}\delta t_{\oplus,1}^{3/5}$.

As we will show in the following section, with $L_{\rm m}\sim
5\times 10^{49}~ {\rm ergs~s^{-1}}$, $\delta t_\oplus\sim 10$ s,
$\Gamma \sim 30$, $\Gamma_{\rm sh}\sim 2$, $\epsilon_{\rm e} \sim
0.5$, and $\epsilon_{\rm B} \sim 0.1$, most of the internal shocks
energy is emitted in the soft X-ray band and the predicted flux
matches the observation of the X-ray flare in GRB 011121.
Alternatively, as shown in Barraud et al. (2005), with
$\epsilon\sim 0.01$ (correspondingly, $L_{\rm m}\sim
10^{51}-10^{52} ~{\rm ergs~s^{-1}}$ and $\Gamma_{\rm sh}\sim
1.06$), $\delta t_\oplus\sim $ a few seconds, and $\Gamma\sim {\rm
a~ few}$ hundreds, X-ray dominated luminosity is expected, too
(Please note that their calculation of the internal shocks
emission is somewhat different from ours; see Barraud et al. 2005
for detail). Therefore, we believe that with proper parameters,
X-ray flares do appear in the late internal shocks scenario.

\subsubsection{The synchrotron radiation of the``late internal shocks"}\label{sec:Radition}
Following Dai \& Lu (2002), the comoving number density of the unshocked outflow
is estimated by $n_{\rm e}\approx L_{\rm m}/(4\pi \Gamma^2 R_{\rm int}^2 m_{\rm p}c^3)$, where $m_{\rm p}$
is the rest mass of proton.  The thermal energy density of the
shocked material is calculated by $e \approx 4\Gamma_{\rm sh} (\Gamma_{\rm
sh}-1)n_{\rm e}m_{\rm p}c^2$ (Blandford \& McKee 1977).
 The intensity of the generated
magnetic field is estimated by $B\approx (8\pi \epsilon_{\rm B}e)^{1/2} \approx 6\times 10^3~{\rm G}~
\epsilon_{\rm B,-1}^{1/2}[\Gamma_{\rm sh}(\Gamma_{\rm
sh}-1)/2]^{1/2} L_{\rm m,49.7}^{1/2}\Gamma_{1.5}^{-1}R_{\rm int,14.5}^{-1}$.

As usual, we assume that in the shock front, the accelerated electrons distribute as $dn_e/d\gamma_{\rm e}\propto
\gamma_{\rm e}^{\rm -p}~~{\rm for~\gamma_{\rm e}>\gamma_{\rm e,m}}$, where $\gamma_{\rm e, m}=\epsilon_{\rm e}
(\Gamma_{\rm sh}-1)[(p-2)m_{\rm p}]/
[(p-1)m_{\rm e}]$ is the minimum Lorentz factor of the shocked electrons (Sari et al. 1998),
and $m_{\rm e}$ is the rest mass of electron. In this section, we take p=2.5.
The observed typical frequency of the
synchrotron radiation reads
\bea
\nu_{\rm m,\oplus} &=& \gamma_{\rm e,m}^2q_e \Gamma B/[2(1+z)\pi m_{\rm e}c]\nonumber\\
&\simeq & 2.7\times 10^{16}~{\rm Hz} ~
\epsilon_{\rm e,-0.3}^2\epsilon_{\rm B,-1}^{1/2}(\Gamma_{\rm sh}-1)^{5/2}(\Gamma_{\rm sh}/2)^{1/2}\nonumber\\
& & L_{\rm m,49.7}^{1/2}\Gamma_{1.5}^{-2}\delta t_{\oplus,1}^{-1},
\label{Eq:nu_m}
\eea
i.e., most of the shock energy is emitted in the soft X-ray band, where $q_e$ is the charge of electron.

The cooling Lorentz factor is estimated by (e.g., Sari et al. 1998) $\gamma_{\rm e,c}
\approx 7.7\times 10^8 (1+z) /(\Gamma B^2\delta t_\oplus)$, and the corresponding cooling frequency reads
\bea
\nu_{\rm c,\oplus} &=& \gamma_{\rm e,c}^2 q_e \Gamma B/[2(1+z)\pi m_{\rm e}c]\nonumber\\
&\simeq & 10^{10}~{\rm Hz}~[1.36/(1+z)]^2
\epsilon_{\rm B,-1}^{-3/2}\Gamma_{1.5}^8\nonumber\\
 & & [\Gamma_{\rm sh}(\Gamma_{\rm sh}-1)/2]^{-3/2} L_{\rm m,49.7}^{-3/2}\delta t_{\oplus,1}.
\label{Eq:nu_c}
\eea
The synchrotron self-absorption frequency is estimated by (Li \& Song 2004)
\be
\nu_{\rm a,\oplus}\simeq 10^{15}~{\rm Hz}~[1.36/(1+z)]^{3/7}L_{\rm
m,49.7}^{2/7}\Gamma_{1.5}^{-5/7}\delta t_{\oplus,1}^{-4/7}B_3^{1/7}.
\ee
The maximum spectral flux of the synchrotron radiation is (e.g., Wijers \& Galama 1999)
$F_{\rm max} \approx 3\sqrt{3}\Phi_{\rm p}(1+z)N_{\rm e}m_{\rm e}c^2\sigma_{\rm T}\Gamma B/(32\pi^2 q_e D_{\rm L}^2)$,
where $N_{\rm e}= L_{\rm m} \delta t/[(1+z)\Gamma m_{\rm p}c^2]=8\times 10^{51}~[1.36/(1+z)]L_{\rm m,49.7}\Gamma_{1.5}^{-1}
\delta t_{\oplus,1}$ is
the number of electrons involved in the emission. $\Phi_{\rm p}$ is a function of $p$, for $p=2.5$, $\Phi_{\rm p}
\approx 0.6$ (Wijers \& Galama 1999). For $\nu_{\rm c, \oplus}<\nu_{\rm a,\oplus}<\nu_{\rm x}<\nu_{\rm m, \oplus} $, the predicted flux
is (e.g., Sari et al. 1998)
\bea
F_{\nu_{\rm x}} &=&
F_{\rm max}(\nu_{\rm m, \oplus}/\nu_{\rm c, \oplus})^{-1/2}(\nu_{\rm x}/\nu_{\rm m, \oplus})^{\rm -p/2}\nonumber\\
&\approx& 2.5~{\rm mJy}~ [\nu_{\rm x}/(2.42\times 10^{17}{\rm Hz})]^{\rm -p/2}\epsilon_{\rm e,-0.3}^{\rm p-1}\epsilon_{\rm B,-1}^{\rm (p-2)/4}\nonumber\\
&& (\Gamma_{\rm sh}/2)^{\rm (p-2)/4} (\Gamma_{\rm sh}-1)^{\rm (5p-6)/4}L_{\rm m,49.7}^{\rm (p+2)/4}\nonumber\\
&& \Gamma_{1.5}^{\rm 2-p}\delta t_{\oplus,1}^{\rm (2-p)/2} D_{\rm L,27.7}^{\rm -2}.
\label{Eq:Flux}
\eea
Taking $Q_{\rm y}=1$ and $\nu_{\rm x}=2.42\times 10^{17}$ Hz, with equation (\ref{Eq:Flux}) we have $F_{\nu_{\rm x}}
\approx 2.5$ mJy,
 which matches the observation of GRB 011121 ($\sim 1$ mJy). The V band flux can be estimated as
($\nu_{\rm c, \oplus}<\nu_{\rm v}<\nu_{\rm a,\oplus}$) $F_{\nu_{\rm v}}\sim
F_{\nu_{\rm max}} \nu_{\rm a,\oplus}^{-3}\nu_{\rm c, \oplus}^{1/2}\nu_{\rm v}^{5/2}\sim 40$ mJy.

What happens after the ``late internal shocks"? Surely, the refreshed relativistic outflow will catch up
with the initial outflow when the latter has swept a large amount of material and got decelerated.
That energy injection would give rise to a flattening (e.g., Rees \&
M\'esz\'aros 1998) or re-brightening signature (e.g., Panaitescu, M\'{e}sz\'{a}ros \& Rees 1998; Kumar \& Piran 2000; Zhang \& M\'esz\'aros 2002), which could
potentially account for the late re-brightening of XRF 050406, the late X-ray re-bursting detected in GRB 050502b and the second/weaker X-ray bump observed in GRB 011121. However, the detailed lightcurve modeling is beyond the scope of this Letter.

\subsubsection{The decline behavior of the flare}
\begin{figure}
\begin{picture}(0,200)
\put(0,0){\includegraphics{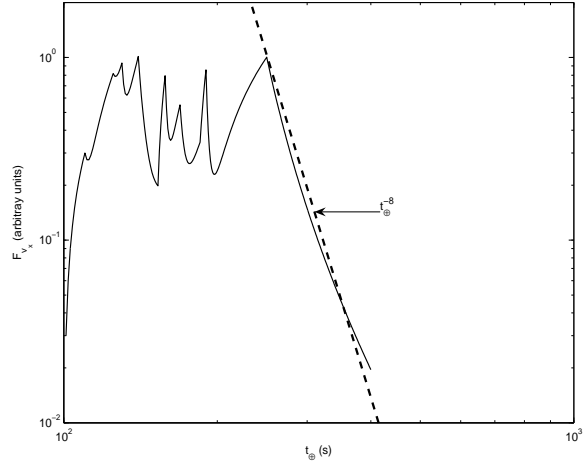}}
\end{picture}
\caption{The X-ray lightcurve of one flare consisting of ten pulses (the solid line, for illustration), each
takes the profile $F_{\nu_{\rm X}}\propto [(t_\oplus-t_{\rm eje})/\delta t_\oplus]$  for
 $t_{\rm eje}<t_\oplus<t_{\rm eje}+\delta t_\oplus$ and
$F_{\nu_{\rm X}}\propto [(t_\oplus-t_{\rm eje})/\delta t_\oplus]^{-3.25}$ for $t_\oplus>t_{\rm eje}+\delta t_\oplus$.
In these pulses ($i=1-10$), the peak of
$F_{\nu_{\rm x},i}$ are taken to be
(0.3,~0.8,~0.6,~0.9,~0.1,~0.7,~0.5,~0.3,~0.7,~1.0), respectively (in arbitrary units); $\delta t_{\oplus, i}$ are taken to be (10,~15,~5,~10,~13,~5,~9,~11,~5,~60) s, respectively;
$t_{{\rm eje},i}$ are taken to be (100,~110,~125,~130,~140,~153,~158,~169,~185,~190) s, respectively.
$t_{\rm e,\oplus}\approx t_{\rm eje,10}+\delta t_{\oplus,10}=250$ s. The dashed line
represents $F_{\nu_{\rm X}}\propto (t_\oplus/t_{\rm e, \oplus})^{-8}$.}
\label{fig:Illustration}
\end{figure}

For observer's frequencies above the cooling frequency, the
curvature effect dominates the temporal behavior of the observed
flux after the central engine shuts down at $t_{\rm e, \oplus}$
(PK00).  The flux declines as ${\sum\limits_{i}} F_{\nu_{\rm x},i}
[(t_\oplus-t_{{\rm eje},i})/\delta t_{\oplus, i}]^{\rm
-(2+p_i/2)}$ (PK00), where $i$ represents the $i$th pulse,
$t_{{\rm eje},i}$ and $t_{\oplus,i}$ are the ejection time and the
variability timescale of the $i$th pulse, respectively.  Such a
decline is much steeper than $(t_\oplus/t_{\rm
e,\oplus})^{-(2+p/2)}$ for $t_{\rm e,\oplus}\gg \max \{\delta
t_{\oplus, i}\}$.
 For example, as shown in Fig. \ref{fig:Illustration}, the decline of the flare is dominated by the
 curvature effect of the last long pulse with $\delta t_\oplus \sim 0.24 t_{\rm e, \oplus}$. A
crude power-law fit to the decline yields $F_{\nu_{\rm X}}\propto
(t_\oplus/t_{\rm e,\oplus})^{-8}$, which is steep enough to match
the sharpest decline detected so far.  In reality, the central
engine does not turn off abruptly. The dimmer and dimmer emission
powered by the weaker and weaker ``late internal shocks" may
dominate over the curvature effect of the early pulses, resulting
in a shallower decay.

\section{Decline behavior of the X-ray flare: constraint on the model}
\label{sec:XRB3} Early X-ray flares have been well detected in GRB 011121, XRF 050406, GRB 050502b and GRB 050730
 (P05; Burrows et al. 2005; Staring et al. 2005). The rise and fall of the first flare (also the dominant one)
 in GRB 011121 are both very steep.
 Similar temporal behavior is evident in other events. The sharp decline of these flares imposes a robust
 constraint on the model, as shown below.

The X-ray flare detected in GRB 011121 appears at $t_{\rm b, \oplus}=239$ s and peaks at $t_{\rm p, \oplus}=270$ s. The burst
is believed to be born in a weak stellar wind. As shown in Fig. \ref{fig:FRX}(b), no flare is
expected in the FS-RS model.  The FS-RS shock model is further disfavored by its shallow decline.
In the late internal shock model, the decline of the flare can be steep enough to
account for the observation (see Fig. \ref{fig:Illustration} for illustration).  Moreover, as shown in
\S\ref{sec:Radition}, with proper parameters the observed flux can be well reproduced. So the
``late internal shock model" is favored. We would like to point out that
{\it the fall of the X-ray flare detected in GRB 011121 is still attributed to the late
internal shocks rather than the curvature effect}. The reason is as follows. Since
 $\delta t_\oplus \leq (t_{\rm p,\oplus}-t_{\rm b,\oplus})=31$ s, the resulted decline $F_{\nu_{\rm x}}\propto [1+(t_\oplus-270)/\delta t_\oplus]^{-3.15}$
is much steeper than the observation $F_{\nu_{\rm x}}\propto
[(t_\oplus-239)/31]^{-1.4}$ (see Fig. 7 of P05) if after $t_{\rm
p,\oplus}$ there are no internal shocks any more.

The X-ray flare detected in XRF 050406 peaks at $t_{\rm p, \oplus}\approx 210$ s and
 declines as $F \propto t_\oplus^{-5.7}$.   The X-ray flare detected in GRB 050502b peaks at
$t_{\rm p, \oplus}\approx 650$ s and declines as $F \propto
t_\oplus^{-7}$. In the X-ray afterglow lightcurve of GRB 050730,
there are three X-ray flares (ranging from 200 s to 800 s after
the trigger of the GRB). A crude fit to the decline of these three
flares results in $F\propto t_\oplus^{-5}$ or steeper. Obviously,
the FS-RS scenario is ruled out by the steep observed decays and
the ``late internal shock model" is favored.
 For the X-ray flare detected in GRB 050502b, the late internal shock model interpretation is further supported by the
 sharp spike detected in 1.0$-$10.0 keV band (Burrows et al. 2005).

\section{Summary \& Discussion}
\label{sec:XRB4}

In this work, we have explored two possible models which might give rise to X-ray flares in GRB afterglows.
One is the external forward-reverse shock model (the ISM case), in which the shock parameters
of forward/reverse shocks are taken to be quite different. {\it The other is the ``late internal shock model", which
requires that a refreshed unsteady relativistic outflow is generated after the prompt $\gamma-$ray emission}
(see  Ramirez-ruiz, Merloni \& Rees [2001] for alternative scenarios), {\it perhaps due to the
fallback accretion onto the central collapsar remnant}. The refreshed outflow may be characterized by
 a low outflow luminosity ($\sim 10^{49}~{\rm ergs~s^{-1}}$), a small bulk Lorentz factor ($\sim 30$), and a long
variability timescale ($\sim 10$ s).
In the external forward-reverse shock model, after the peak
 of the reverse shock emission ($t_{\rm p,\oplus}=t_\times$), the flux
can not decline more sharply than $(t_\oplus/t_{\rm p,\oplus})^{\rm -(2+p/2)}$ (see Fig. \ref{fig:FRX} for illustration). In
the ``late internal shock model", the decline can be much steeper than $(t_\oplus/t_{\rm e,\oplus})^{\rm -(2+p/2)}$
 if the central engine shuts down after $t_{\rm e,\oplus}$ and the
longest variability timescale of the X-ray flare is much shorter than $t_{\rm e,\oplus}$
(see Fig. \ref{fig:Illustration} for illustration).

For the X-ray flares detected in GRB 011121, XRF 050406, GRB 050502b and GRB 050730, the external forward-reverse shock model
is ruled out directly by its shallow temporal decay. For the same reason,
 other possible external models
(i.e., the model related to the external forward shock), including
the density jump model, the two-components jet model, the patch jet model as well as the energy injection model are
ruled out too (Zhang et al. 2005). Thus, the ``late internal shock model" is found to be favored. In this model, the optical
emission may be suppressed due to strong synchrotron-self-absorption.
 But in the ultraviolet band, the radiation could be quite strong. Large amount of neutral gas would be ionized, as detected in GRB
050502b and GRB 050730 (Burrows et al. 2005; Starling et al. 2005).

{\it Very early X-ray flares are well detected  both in long
GRBs and in XRFs,  which strengthens the correlation of these two phenomena}, though the nature of
XRFs is still unclear (Barraud et al. 2005 and the references therein).

Finally, we suggest that the early X-ray light curve of some GRBs may be a superposition of the emission powered by
the long activity of the central engine and the emission of
the external forward shock. As a consequence, the X-ray temporal behavior may be quite different from that of the long
wavelength emission (UV/Optical ones). This prediction can be tested by the UVOT
and XRT on board {\em Swift} observatory directly in the near future.

\section*{Acknowledgments}
We thank Bing Zhang and E. W. Liang for
informing us Piro et al.'s  paper on GRB 011121 at the end of Feb, and G. F. Jiang, L. J. Gou, Z. Li and H. T. Ma
for kind help. We also appreciate the referees for their helpful comments and the third referee for
her/his great help. This work is supported by the National Natural Science Foundation
(grants 10225314 and 10233010) of China, and the National 973
Project on Fundamental Researches of China (NKBRSF G19990754).


\begin{appendix}
\section{Derivation of the curvature effect}
Assuming a shell is ejected at $t_{\rm
eje}$ and moves with a constant Lorentz factor $\Gamma$ (the corresponding velocity is $V$, in unit of the
speed of light),
the electrons are shock-accelerated at $R<R_{\rm cro}$ and
cools rapidly, where $R_{\rm cro}$ is the radius after which there is no newly relativistic electrons injected.
 For $R>R_{\rm cro}$, the radiation nearly cuts off as long as the observer
frequency $\nu_\oplus$ is above the cooling frequency of the electrons.  For $t_\oplus>t_0\equiv t_{\rm eje}+(1+z)R_{\rm
cro}/(2\Gamma^2 c)$, the flux received from the shell is given by
\begin{equation}
F_{\rm \nu_\oplus}(t_\oplus)\propto \int_{\theta_t}^{\theta_{\rm j}}
{ {\cal S}_{\rm \nu'} {\rm sin}\theta d\theta \over \Gamma^3(1-V {\rm cos}\theta)^3},
\label{Eq:curv1}
\end{equation}
where $\theta_t$ satisfies $R_{\rm cro}\approx c(t_\oplus-t_{\rm eje})/[(1+z)(1-V \cos \theta_t)]$, $\theta_{\rm
j}$ is the half opening angle of the shell, $\nu'=\Gamma(1-V {\rm cos}\theta)\nu_{\oplus}$, ${\cal S}_{\nu'}\propto
(R/R_{\rm cro})^{\rm k}\nu'^{\rm -\beta}$ is the specific spectrum of the radiation in unit of solid angle.   On the
``equal arriving time surface" $R(\theta)=c(t_\oplus-t_{\rm
eje})/[(1+z)(1-V \cos \theta)]$ (Rees 1966, Nature, 211, 468), the magnetic field, the typical emission frequency of the
electrons, and the number of electrons involved in the radiation are all functions of $R$, that's why we take into account
 the term $(R/R_{\rm cro})^{\rm k}$ in calculating ${\cal S}_{\nu'}$.

Equation (\ref{Eq:curv1}) yields
\begin{eqnarray}
F_{\nu(t_\oplus)} &\propto & \int^{\theta_j}_{\theta_t} (1-V \cos \theta_t)^{\rm k}(1-V \cos \theta)^{\rm -(3+\beta +k)}\sin\theta d \theta \nonumber\\
& \approx & {1\over 2+\beta +k}(1-V \cos \theta_t)^{\rm -(2+\beta)}.
\label{Eq:curv2}
\end{eqnarray}
On the other hand, $F_{\nu(t_0)} \propto {1\over 2+\beta +k}(1-V)^{\rm -(2+\beta)}$. For $t_0<t_\oplus<t_j$, we have
\begin{equation}
F_{\nu(t_\oplus)}=F_{\nu(t_0)}[{1-V\cos \theta_t \over 1-V}]^{\rm -(2+\beta)}\propto ({t_\oplus-t_{\rm eje}\over t_0-t_{\rm eje}})^{\rm -(2+\beta)},
\label{Eq:curv3}
\end{equation}
which coincides with that presented in KP00 but derived in a different way.
\footnote{The derivation of the curvature effect has not been presented in the paper to match the page limit.}
\end{appendix}

\end{document}